\documentclass[11pt]{article}
\pdfoutput=1
\newcommand{\Comment}[1]{{}}
\usepackage{amsfonts,amsthm,amsmath,amssymb,slashed}
\usepackage[textwidth = 430 pt, textheight = 630 pt]{geometry}
\usepackage{color}
\usepackage{booktabs}
\usepackage{subcaption}
\usepackage[export]{adjustbox}

\Comment{\usepackage{color}
\definecolor{MyDarkBlue}{rgb}{0.15,0.15,0.45}
\usepackage[linktocpage=true]{hyperref}
\hypersetup{
colorlinks=true,
citecolor=MyDarkBlue,
linkcolor=MyDarkBlue,
urlcolor=MyDarkBlue,
pdfauthor={Horatiu Nastase and Caio Luiz Tiedt},
pdftitle={Susceptibilities from holographic transport with topological term},
pdfsubject={hep-th}
}
\usepackage[utf8]{inputenc}
\usepackage[numbers,sort&compress]{natbib}
\usepackage{hypernat}}
\usepackage{graphicx}
\usepackage{cite}
\usepackage[colorlinks=true]{hyperref}
\hypersetup{
  allcolors = blue,
}

\newcommand\ignore[1]{}
\def\one{{\,\hbox{1\kern-.8mm l}}}

\def\a{\alpha}

\def\d{\partial}

\newcommand{\Cset}{{\,\,{{{^{_{\pmb{\mid}}}}\kern-.45em{\mathrm C}}}}}

\newcommand{\be}{\begin{equation}}
\newcommand{\bea}{\begin{eqnarray}}

\newcommand{\ee}{\end{equation}}
\newcommand{\eea}{\end{eqnarray}}

\begin{document}

\renewcommand{\thefootnote}{\fnsymbol{footnote}}

\makeatletter
\@addtoreset{equation}{section}
\makeatother
\renewcommand{\theequation}{\thesection.\arabic{equation}}

\rightline{}
\rightline{}




\begin{center}
{\LARGE \bf{\sc Susceptibilities from holographic transport with topological term}}
\end{center}
 \vspace{1truecm}
\thispagestyle{empty} \centerline{
{\large \bf {\sc Horatiu Nastase${}^{a}$}}\footnote{E-mail address: \Comment{\href{mailto:horatiu.nastase@unesp.br}}{\tt horatiu.nastase@unesp.br}}
{\bf{\sc and}}      
{\large \bf {\sc Caio Luiz Tiedt${}^{b}$}}\footnote{E-mail address: \Comment{\href{mailto:caio.tiedt@if.usp.br}}{\tt caio.tiedt@if.usp.br}}                                           }

\vspace{.5cm}


\centerline{{\it ${}^a$Instituto de F\'{i}sica Te\'{o}rica, UNESP-Universidade Estadual Paulista}}
\centerline{{\it R. Dr. Bento T. Ferraz 271, Bl. II, Sao Paulo 01140-070, SP, Brazil}}
\vspace{.3cm}
\centerline{{\it ${}^b$Instituto de F\'{i}sica, Universidade de S\~ao Paulo,}}
\centerline{{\it Rua do Mat\~ao, 1371, 05508-090 S\~ao Paulo, SP, Brazil}}

\vspace{1truecm}

\thispagestyle{empty}

\centerline{\sc Abstract}

\vspace{.4truecm}

\begin{center}
\begin{minipage}[c]{380pt}
{\noindent In this  note we calculate susceptibilities, as derivatives of a thermodynamical 
potential, for the general perturbative holographic set-up for transport with magnetic field, charge density and topological term, 
and compare with the quantities obtained in the standard $AdS_4$ dyonic black hole analysis of Hartnoll and Kovtun. We find that 
the results do not match, despite previous expectations.
}
\end{minipage}
\end{center}

\vspace{.5cm}

\setcounter{page}{0}
\setcounter{tocdepth}{2}

\newpage

\tableofcontents
\renewcommand{\thefootnote}{\arabic{footnote}}
\setcounter{footnote}{0}

\linespread{1.1}
\parskip 4pt



\section{Introduction}

The AdS/CFT correspondence \cite{Maldacena:1997re} (see \cite{Nastase:2015wjb,Ammon:2015wua} for a review)
usually relates strongly coupled field theory to weakly coupled string theory in its classical supergravity limit, with "top-down" models,
derived from systems of branes in a decoupling limit. 
Common applications to condensed matter, AdS/CMT (see \cite{Nastase:2017cxp}
for a review), are usually phenomenological, "bottom-up" constructions. That applies in particular to models of transport 
in condensed matter systems. 

However, there are a few examples of top-down models as well, most notably the ABJM vs. $AdS_4\times \mathbb{CP}^3$
correspondence \cite{Aharony:2008ug}, which has been used as a sort of a prototype for transport in strongly coupled 
2+1 dimensional condensed matter systems. Of course, it is not a top-down model in the sense that there is no {\em derived} 
relation of the ABJM model to any condensed matter system (unlike supersymmetric $SU(N)$ gauge theories in 3+1 
dimensions, thought of as an extension of the gluon theory for QCD), only a phenomenological one: it gives similar physics. 
But the holographic map is derived. At nonzero temperature, the dyonic black hole in $AdS_4$ has been used as a model 
for 2+1 dimensional transport in the presence of a magnetic field \cite{Hartnoll:2007ai,Hartnoll:2007ih}. One can calculate 
thermodynamic quantities, and transport from fluctuations around the dyonic background. Note that these were extended
to the presence of a topological term in the action in \cite{Nastase:2022etj}.

However, the generic transport is necessarily obtained from a background obtained by adding perturbations at infinity
(and perhaps the horizon of the black hole), so that the full background solution is not known, following the 
method in 
\cite{Blake:2015ina,Donos:2015bxe,Donos:2014cya,Banks:2015wha,Donos:2014uba,Donos:2017mhp,Erdmenger:2016wyp}. 
One rather generic case was 
considered in \cite{Alejo:2019utd}. In \cite{Melnikov:2020ktj}, the Wiedemann-Franz law was obtained by a combination 
of the two methods. In particular, the matrix of susceptibilities $\chi_s$, calculated as the second order derivatives
of the thermodynamic potential in the dyonic black hole background, and was related via the matrix of diffusivities $D$
to the matrix of conductivities (as expected from the general theory of the hydrodynamic limit), for which the results in
the perturbative background from \cite{Alejo:2019utd}. 

But that implies the assumption that dyonic black hole background of \cite{Hartnoll:2007ai,Hartnoll:2007ih} and the perturbative 
one of \cite{Alejo:2019utd} give the same thermodynamics, which is not obvious. Therefore in this paper we investigate the 
possibility of these two results giving the same answer. This has implications beyond the specific case considered here, 
as it measures the correctness of importing results from a top-down construction to a bottom-up one, or vice versa.

The paper is organized as follows. In section 2 we consider the perturbative model with topological term, 
but only $B,B_1$ external fields, and calculating the thermodynamics, the  magnetizations and the susceptibilities 
with this simplified version of the fluctuations.
In section 3, we calculate the transport coefficients for a more general version of the model, with $E$ and $\xi=(\nabla T)/T$
as external fields as well.
In section 4, we calculate the susceptibilities for this general case, and compare with the $AdS_4$ dyonic black hole results. 
We conclude in section 5.

\section{AdS/CMT perturbative model and boundary conditions at the black hole horizon}

For the gravitational theory with $AdS_4$ asymptotics, 
we consider a $3+1$ dimensional Einstein-Hilbert action (where the spacetime is described in coordinates $t,r,x,y$) 
for gravity with a scalar dilaton $\phi$, an $U(1)$ gauge field with a Maxwell term with kinetic function $Z(\phi )$ 
and a topological term with function $W(\phi )$. We also add two linear axions $\xi_1$ and $\xi_2$ with action 
proportional to a function $\Phi(\phi)$, in order to break translational invariance, as needed for transport. 
The action is thus given by:

\begin{equation}
\begin{split}
    I  & =  \int dx^4 \sqrt{-g}\left[
    \frac{1}{16 \pi  G_N}\left(R-V(\phi)-\frac{1}{2} (\partial^\mu\phi ) (\partial_\mu\phi )-\frac{1}{2} ((\partial \chi_1)^2
    +(\partial \chi_2)^2) \Phi (\phi)\right)
    \right.\\
    & \hspace{250pt}  \left.
    - \frac{F_{\mu\nu}F^{\mu\nu} Z(\phi)}{4 g_4^2}
    - F_{\mu\nu}\tilde{F}^{\mu\nu} W(\phi   ) 
    \right]\;,    
    \end{split}
\end{equation}
where, as usual, $ F_{\mu\nu} = \partial_\mu A_\nu - \partial_\nu A_\mu$ and $
    \tilde{F}_{\mu\nu} = \frac{\epsilon^{\mu\nu\delta\rho}}{2\sqrt{-g}}F_{\delta\rho}$.

This model without the topological term 
has been studied by \cite{Blake:2015ina} and in \cite{Alejo:2019utd} the authors added the topological term to it.

The equations of motion for this model are: -for the metric field:
\begin{equation}
    R_{\mu\nu}  = \frac{1}{2}\partial_\mu\phi\partial_\nu\phi + \frac{1}{2}V(\phi) + \frac{16 \pi G_N}{4g_4^2} 
    \left( 2 F_{\mu\sigma}F_\nu^\sigma - \frac{1}{2}g_{\mu\nu}F_{\sigma\rho}F^{\sigma\rho}\right),
\end{equation}
-for the connection $A_\mu$:
\begin{equation}
    \frac{1}{\sqrt{-g}} \partial_\mu \sqrt{-g}\left(F^{\mu\nu} + W(\phi)\tilde{F}^{\mu\nu} \right) = 0,
\end{equation}
-for the axion fields:
\begin{equation}
    \Phi(\phi ) \partial_\mu \partial^\mu \chi_i + \partial_\mu \chi_i\partial^\mu \phi \Phi'(\phi) = 0
\end{equation}
-and finally for the dilaton field:
\begin{equation}
    (\partial_\mu \phi)^2 - V'(\phi) - \frac{1}{2}((\partial_\mu \chi_1)^2 -(\partial_\mu \chi_2)^2) \Phi'(\phi) 
    = 16 \pi G_N\left(     \frac{F_{\mu\nu}F^{\mu\nu} Z'(\phi)}{4 g_4^2}
    + F_{\mu\nu}\tilde{F}^{\mu\nu} W'(\phi   ) \right).
\end{equation}

The axion fields, have background solutions
\be
        \chi_1  = k_1 x\;,\;\;\;
        \chi_2  = k_2 x\;,
\ee
that break translational invariance, as we need, and $k_1\neq k_2$ would also break isotropy. 

The background solution for this model is given by an asymptotically $AdS_4$ metric
\begin{equation}
    ds^2 = -U(r) dt^2 + \frac{1}{U(r)} dr^2 + e^{2V(r)}(dx^2+dy^2),
\end{equation}
while the $U(1)$ gauge field is such that the boundary field theory has 
a magnetic field $B$ and an electric field defined by $a(r)$, so is 
\begin{equation}
    A = a(r)dt - B y dx.
\end{equation}

In order for us to have a holographic dual (with $AdS_4$ asymptotics), we require that the scalar potential $V(\phi)$ satisfies
\begin{equation}
    V(0) = -\frac{6}{L^2}, \hspace{25 pt} V'(0) = 0.    
\end{equation}

\subsection{Thermodynamics and magnetization}

In the above background, we want to study electrical and thermal transport in the presence of a magnetic field. 
We consider the {\em Euclidean} action in the bulk, in the absence of axion perturbations, as
\begin{equation}
    S_E = \int d^4x\sqrt{g} \left( \frac{1}{16\pi G_N}\left(R + \frac{1}{2}(\partial \phi )^2 
    +V(\phi)\right) +\frac{Z(\phi)}{4g_4^2}F_{\mu\nu}F^{\mu\nu} - W(\phi) F_{\mu\nu}\tilde{F}^{\mu\nu} \right).
\end{equation}

In this case,  the response of the euclidean action with the change in the magnetic field gives the magnetization density,
\begin{equation}
    M = -\frac{1}{\text{Vol}}\frac{\partial S_E}{\partial B}.
\end{equation}

We also need to consider the response of the action with respect to a fluctuations in the metric of the type 
$\delta g_{tx}=-B_1 y$, which gives the energy magnetization density,
\begin{equation}
    M_E = -\lim_{B_1\rightarrow 0}\frac{1}{\text{Vol}}\frac{\partial S_E}{\partial B_1}.
\end{equation}

These two affect the background solutions by adding a term to $A_x$ and a non-diagonal term to the metric,
\begin{align}
    A = & a(r)dt + (-B_1 + (a(r)-\mu)B_1y) dx ,\\
    ds^2 = & -U(r)(dt+B_1ydx)^2 + \frac{dr^2}{U(r)} +e^{2V(r)}(dx^2+dy^2).
\end{align}

Then on-shell, the gauge terms equal 
\begin{align}
    F_{\mu\nu}F^{\mu\nu} & = 2 E^{-4 V(r)} (B + B_1 \mu - B_1 a(r))^2 - 2 a'(r)^2\\
    F_{\mu\nu}\tilde{F}^{\mu\nu} & = 4 e^{-2 V(r)} a'(r) (-B_1 a(r)+B+B_1 \mu)\;,
\end{align}
and the Ricci scalar is given by
\begin{equation}
    R = U(r) \left(\frac{1}{2} B_1^2 e^{-4 V(r)}-6 V'(r)^2-4 V''(r)\right)-4 U'(r) V'(r)-U''(r).
\end{equation}

Taking the derivatives 
\be\label{eqderiva1}
    \frac{-1}{\text{Vol}}\frac{\partial S_E}{\partial B} = \int_{r_h}^\Lambda dr
    \left(\frac{B+B_1\mu-B_1a(r)}{g_4^2}e^{-2V(r)} Z(\phi) - 4W(\phi)a'(r) \right) 
\ee
\bea
    \frac{-1}{\text{Vol}}\frac{\partial S_E}{\partial B_1} &=& \int_{r_h}^\Lambda dr\left[\frac{B_1 e^{-2V(r)} U(r)}{16\pi G_N}\right.\cr
    &&\left.+\left(\frac{B+B_1\mu-B_1a(r)}{g_4^2}e^{-2V(r)} Z(\phi) - 4W(\phi)a'(r) \right)(\mu-a(r))\right] \;\label{eqderiva2}
\eea
and then the limit of $B_1$ going to zero, we get the magnetization densities,
\begin{equation}\label{eq:magequ}
    M = \int_{r_h}^\Lambda dr\left(\frac{B e^{-2V(r)} Z(\phi)}{g_4^2} - 4W(\phi)a'(r) \right) 
\end{equation}
\begin{equation}
    M_E = \int_{r_h}^\Lambda dr\left(\frac{B e^{-2V(r)} Z(\phi)}{g_4^2} - 4W(\phi)a'(r) \right)(\mu-a(r)).
\end{equation}

We can also define the heat magnetization density as $M_Q = M_E - \mu M$, giving
\begin{equation}
    M_Q = - \int_{r_h}^\Lambda dr \left(\frac{B e^{-2V(r)} Z(\phi)}{g_4^2} - 4W(\phi)a'(r) \right)a(r).
\end{equation}

\subsection{Susceptibilities}

We define the susceptibilities, as usual, as the second derivatives of the thermodynamical potential, which in holography 
equals the Euclidean action, which is {\em a priori} a function of $B,B_1,\mu$ and $T$, and in the most general 
case to be considered in the next section, also of $E$ and $\xi$, $S_E(B,B_1,\mu,T;E,\xi)$. 
In this case, the magnetization susceptibility is the derivative of the magnetization with respect to $B$,
\begin{equation}
    \chi_{BB} = \frac{\partial}{\partial B}\left. \left( -\frac{1}{\text{Vol}}\frac{\partial S_E}{\partial B}\right)\right|_{B_1,\mu,T}\;,
\end{equation}
and more generally, for replacing any $B$ with a $B_1$, so 
\begin{equation}
    \chi_{B_i B_j}  = \frac{\partial }{\partial B_i}\left.\left( -\frac{1}{\text{Vol}}\frac{\partial S_E}{\partial B_j}\right)\right|_{B_k,\mu,T}.
\end{equation}

Then, by taking the derivative of equations (\ref{eqderiva1}-\ref{eqderiva2}), we get
\begin{equation}
\begin{split}
    \chi_{BB} & = \int_{r_h}^\Lambda \frac{e^{-2V(r)} Z(\phi)}{g_4^2}dr, \\
    \chi_{BB_1} & = \int_{r_h}^\Lambda \frac{e^{-2V(r)} Z(\phi)}{g_4^2}(\mu-a(r))dr, \\
    \chi_{B_1B_1} & = \int_{r_h}^\Lambda \frac{e^{-2V(r)} U(r)}{16\pi G_n}+\frac{e^{-2V(r)} Z(\phi)}{g_4^2}(\mu-a(r))^2dr. \\
\end{split}    
\end{equation}

For completeness, we can calculate also the derivatives of the heat magnetizations $M_Q$, 
\begin{equation}
    \left.\frac{\d M_Q}{\d B}\right|_{\mu, T} = -\int_{r_h}^\Lambda \frac{e^{-2V(r)} Z(\phi)}{g_4^2}(a(r))dr
\end{equation}
and
\begin{equation}
    \left.\frac{\d M_Q}{\d B_1}\right|_{\mu, T} 
    = -\int_{r_h}^\Lambda \frac{e^{-2V(r)} U(r)}{16\pi G_n}+\frac{e^{-2V(r)} Z(\phi)}{g_4^2}(-\mu a(r)+a(r)^2)dr.
\end{equation}

Here we see that no terms proportional to $W(\phi)$ appear in the susceptibilities $\chi_{B_iB_j}$.

\section{Ansatz and transport coefficients}\label{sec:Transport}

We are interested in calculating the susceptibilities for the model with general perturbations, 
using the ansatz from \cite{Alejo:2019utd}. In this section we review the calculation of the transport coefficients 
in \cite{Alejo:2019utd}, since the results are going to be used later.

The source of fluctuations are the same as in the previous section at $B_1=0$, a magnetic field $A_x^{(0)} = -By$, 
but now we also consider a nonzero
electric field $E_x=E$ and thermal gradient $\frac{1}{T}\nabla_x T=\xi$. We also add general fluctuations 
for all fields depending on the sources from the Einstein equations of motions, $\delta h_{\mu\nu}$ 
for the metric, $\delta A_\mu$ for the gauge field and $\delta  \chi_i$ for the axion fields.

The resulting fields with all their fluctuations are: 
-the metric field,
\begin{equation}\label{eq:ansatzIni}
g_{\mu\nu} = 
\left(
\begin{array}{cccc}
 -U(r) & 0 & \epsilon  (\delta h_{tx} e^{2 V(r)}-\xi  t  U(r) )& \epsilon  \delta h_{ty} e^{2 V(r)} \\
 
 0 & \frac{1}{U(r)} & \epsilon  \delta h_{rx} e^{2 V(r)} & \epsilon  \delta h_{ry} e^{2 V(r)} \\
 
 \epsilon  (\delta h_{tx} e^{2 V(r)}-\xi  t  U(r)) & \epsilon  \delta h_{rx} e^{2 V(r)} & e^{2 V(r)} & 0 \\
 
 \epsilon  \delta h_{ty} e^{2 V(r)} & \epsilon  \delta h_{ry} e^{2 V(r)} & 0 & e^{2 V(r)} \\
\end{array}
\right)
\end{equation}
-the gauge field,
\begin{equation}
\begin{split}
    A_t = & a(r)\\
    A_x = & -By + t \epsilon (\xi a(r) - E ) + \epsilon \delta A_x\\
    A_y = & \epsilon \delta A_y
\end{split}    
\end{equation}
-and the axion fields
\begin{equation}
     \chi_1(r) = k_1 x+\epsilon  \delta \chi_1
\end{equation}
\begin{equation}\label{eq:ansatzFim}
    \chi_2(r) = k_2 y+\epsilon  \delta \chi_2.
\end{equation}

Here, $\epsilon$ is added as a mathematical tool in order to account for the order in the fluctuations, since we are 
considering $B$, $E$ and $\xi$ small.

\subsection{Maxwell's equations of motion}

The gauge equations of motion are given by
\begin{equation}
    \frac{1}{\sqrt{-g}} \partial_\mu \sqrt{-g}\left(F^{\mu\nu} + W\tilde{F}^{\mu\nu} \right) = 0.
\end{equation}

In the equation for $\nu = x$ only the $\mu=r$ term survives
\begin{equation}
    \partial_r \left[\sqrt{-g}\left(F^{r x} + W\tilde{F}^{rx} \right) \right]= 0.
\end{equation}
and in the equation for $\nu = y$ we have an extra term that also survives
\begin{equation}
    \partial_r \left[\sqrt{-g}\left(F^{r y} + W\tilde{F}^{ry} \right)\right] =   
    -4 \xi  a'(r) W(\phi )\epsilon-\frac{B \xi  Z(\phi )}{g_4^2 }\sqrt{e^{-2 V(r)}}\epsilon.
\end{equation}
Note that the second term in the equation above is the same as the integrand times $\xi$ 
of the magnetization found in (\ref{eq:magequ}).

As explained in \cite{Alejo:2019utd}, in order to do a calculation without having the full solution, we can take advantage 
of the fact that there are generalized currents that are $r$-independent, $\mathcal{J}_i$, following the general 
idea of the membrane paradigm in the form of Iqbal and Liu \cite{Iqbal:2008}. 
We can define these currents for the model modified by the magnetization term as
\begin{equation}
    \begin{split}\label{eq:defCurrent}
        \mathcal{J}^x = & \sqrt{-g}\left(F^{r x} + W\tilde{F}^{rx} \right) \\
        \mathcal{J}^y = & \sqrt{-g}\left(F^{r y} + W\tilde{F}^{ry} \right) - \epsilon \xi  M(r). \\
    \end{split}
\end{equation}

Note that we do in fact have $\partial_r \mathcal{J}^i = 0$. On-shell, we have 
\begin{equation}
    \begin{split}
        \mathcal{J}^x = & \frac{\epsilon  e^{4 V(r)} Z(\phi) \left(\delta h_{tx} a'(r)+U(r) e^{-2 V(r)} 
        \left(B \delta h_{ry}+\delta A_x'\right)\right)}{g_4^2} \\
        \mathcal{J}^y = & \frac{\epsilon  e^{4 V(r)} Z(\phi) \left(\delta h_{ty} a'(r)+U(r) e^{-2 V(r)} 
        \left(\delta A_y'-B \delta h_{rx}\right)\right)}{g_4^2}+\\
        & \hspace{110pt}+ 4
   \epsilon  e^{4 V(r)} W(\phi) (\xi  a(r)-E)- \epsilon\xi  M(r). \\
    \end{split}
\end{equation}

The currents are going to be useful for us because they don't depend on the coordinate $r$, 
and thus we can relate the fields for any $r$ to values at the horizon or the boundary:
\begin{equation}
    \mathcal{J}^i (r) = \mathcal{J}^i(r_h) = \lim_{r\rightarrow \infty}\mathcal{J}^i(r).
\end{equation}

At the horizon we have that the magnetization vanishes, $M(r_h) = 0$, directly from the result in (\ref{eq:magequ}). 
Also, we can impose regularity conditions near the horizon \cite{Blake:2015ina},
\begin{align}\label{eq:regularityCondi1}
    \delta A_x & = -\frac{E \ln (r-r_h)}{4 \pi  T} + \mathcal{O}(r-r_h)\\
    \delta A_y & =  \mathcal{O}(r-r_h)\\
    \label{eq:regularityCondi2}
    \delta h_{rx} & = \frac{\delta h_{tx}}{U(r)}+\frac{\xi  e^{-2 V(r)} \log (r-r_h)}{4 \pi  T}+\mathcal{O}(r-r_h)\\
    \delta h_{ry} & = \frac{\delta h_{ty}}{U(r)}+\mathcal{O}(r-r_h)\\
    \delta \chi_i & = \mathcal{O}(r-r_h).
\end{align}

We then obtain that at the horizon the usual currents $J^i$ equal the generalized currents $\mathcal{J}^i$, and equal
\begin{align}\label{eq:current1}
     J^x=\mathcal{J}_x (r_h) & = \lim_{r\rightarrow r_h} \frac{e^{2 V(r)} Z(\phi) \left(\delta h_{tx} e^{2 V(r)} a'(r)+B \delta h_{ty}-E\right)}{g_4^2} 
     \\\label{eq:current2}
     J^y=\mathcal{J}_y (r_h) & = \lim_{r\rightarrow r_h} \frac{e^{2 V(r)} Z(\phi) \left(\delta h_{ty} e^{2 V(r)} a'(r)-B \delta h_{tx}\right)}{g_4^2}+4 W(\phi) (\xi  a(r)-E). \\
\end{align}

We still need to deal with $\delta h_{ti}$, appearing in the above formulas, and for that we must use the gravity equations of motion.

\subsection{Einstein's equations of motion}

The equations of motion for gravity are
\begin{equation}\label{eomGravity}
    R_{\mu\nu}  = \frac{1}{2}\partial_\mu\phi\partial_\nu\phi + \frac{1}{2}V(\phi) + \frac{16 \pi G_N}{4g_4^2} 
    \left( 2 F_{\mu\sigma}F_\nu^\sigma - \frac{1}{2}g_{\mu\nu}F_{\sigma\rho}F^{\sigma\rho}\right).
\end{equation}    

We calculate them on-shell at linear level in $\epsilon$. The derivation of the formulas can get involved, so, 
since we are interested in the result near horizon, we can expand the background field near this region, 
\begin{equation}
\begin{split}
    a(r) & = a_h (r-r_h) + ...\\
    V(r) & = V_h  + ...\\
    \phi & = \phi_h + ...
\end{split}
\end{equation}

Using this, for $\mu\nu = ty$, we end up with  the Einstein equation
\begin{equation}
\begin{split}
    \frac{1}{2}   U(r) e^{2 V(r)} \left(\delta h_{ty}''+4 \delta h_{ty}' V'(r)\right) 
    -B^2   \delta h_{ty} e^{-2 V(r)} Z(\phi) +\\
    \quad +
    \frac{1}{4}   \left(k_1^2+k_2^2\right) \delta h_{ty} e^{2 V(r)} \Phi (\phi)
    -2 B   h_{rx} U(r) a'(r) Z(\phi)\\
    \quad=
    -2   U(r) a'(r) \delta A_{y}' Z(\phi) 
    +2 B   e^{-2 V(r)} Z(\phi ) ( \xi  a(r)- E).
\end{split}    
\end{equation}

We can rewrite the above expression so we get a more familiar result\cite{Blake:2015ina}
\begin{equation}
\begin{split}
    U ( e^{4V} \delta h_{ty}' )' - \left(\frac{\kappa}{g_4^2} B^2 Z + \frac{1}{2}(k_1^2+k_2^2) e^{4V} \Phi \right) 
    \delta h_{ty} - \frac{2\kappa}{g_4^2} U B Z e^{2V} a' h_{rx} = \\ = - \frac{2\kappa}{g_4^2} U Z e^{2V} 
    a' \delta A_y' + \frac{2\kappa}{g_4^2} B Z (\xi a - E).
\end{split}    
\end{equation}

A similar expression can be found for $\mu\nu = tx$,
\begin{equation}
\begin{split}
    U ( e^{4V} \delta h_{tx}' )' - \left(\frac{\kappa}{g_4^2} B^2 Z + \frac{1}{2}(k_1^2+k_2^2) e^{4V} \Phi \right) 
    \delta h_{tx} - \frac{2\kappa}{g_4^2} U B Z e^{2V} a' h_{ry} = \\ = - \frac{2\kappa}{g_4^2} U Z e^{2V} a' \delta A_x'.
\end{split}    
\end{equation}

We need impose the regularity conditions (\ref{eq:regularityCondi1} - \ref{eq:regularityCondi2}), and another 
expansion near the horizon for the function $U(r)$,
\begin{equation}
    U(r) = (r-r_h) U'(r_h) + ...,
\end{equation}
where the coefficient in the expansion is given, as usual, by the temperature
\begin{equation}
    U'(r_h) = 4\pi T.
\end{equation}

Note that 
\begin{equation}
    \delta A_x' = -\frac{E}{4\pi T} \frac{1}{r-r_h} = -\frac{E}{U}.
\end{equation}

Then we have
\begin{align}
    \left( \frac{\kappa}{g_4^2} ZB^2 + \frac{1}{2}e^{2V}(k_1^2+k_2^2)\Phi \right) \delta h_{tx} 
    - \frac{2\kappa}{g_4^2} ZBe^{2V}a_h \delta h_{ty} & = - \frac{2\kappa}{g_4^2}Ze^{2V} a_h E + e^{2V}4\pi T\xi\nonumber
    \\
    \left( \frac{\kappa}{g_4^2} ZB^2 + \frac{1}{2}e^{2V}(k_1^2+k_2^2)\Phi \right) \delta h_{ty} 
    - \frac{2\kappa}{g_4^2} ZBe^{2V}a_h \delta h_{tx} & = - \frac{2\kappa}{g_4^2}Z B E\;,
\end{align}
and we can solve for $\delta h_{tx}$ and $\delta h_{ty}$ in terms of $\xi, E,B$. 

With this result, we can rewrite the currents (\ref{eq:current1}-\ref{eq:current2}) and then equate with the general formula 
for transport
\begin{equation}\label{eq:OhmLaw}
    J_i = \sigma_{xi} E - \alpha_{xi} T\xi\;,
\end{equation}
and thus we can identify the thermoeletric transport coefficients, obtaining
(when comparing with \cite{Alejo:2019utd} note that here we have considered the more general case with $k_1\neq k_2$)
\begin{align}
    \sigma_{xx} & = \left.\frac{1}{2}\frac{e^{2V}(k_1^2+k_2^2)\Phi(2\kappa_4^2g_4^4\rho^2 
    + 2\kappa_4^2B^2Z^2+g^2_4Ze^{2V}(k_1^2+k_2^2)\Phi/2)}{4\kappa_4^4g_4^4B^2\rho^2
    +(2\kappa_4^2B^2Z+g^2_4e^{2V}(k_1^2+k_2^2)\Phi/2)^2} \right|_{r_h}\\ 
    \sigma_{xy} & = \left.4\kappa_4^2B\rho\frac{\kappa_4^2g_4^4\rho^2 + \kappa_4^2B^2Z^2
    +g^2_4Ze^{2V}(k_1^2+k_2^2)\Phi/2}{4\kappa_4^4g_4^4B^2\rho^2+(2\kappa_4^2B^2Z
    +g^2_4e^{2V}(k_1^2+k_2^2)\Phi/2)^2} -4W \right|_{r_h}\\ 
    \alpha_{xx} & = \left.\frac{2\kappa_4^2g_4^4s\rho e^{2V}(k_1^2+k_2^2)\Phi/2} 
    {4\kappa_4^4g_4^4B^2\rho^2+(2\kappa_4^2B^2Z+g^2_4e^{2V}(k_1^2+k_2^2)\Phi/2)^2}  \right|_{r_h}\\ 
    \alpha_{xy} & = \left.2\kappa_4^2sB\frac{2\kappa_4^2g_4^4\rho^2 + 2\kappa_4^2B^2Z^2+g^2_4Ze^{2V}(k_1^2+k_2^2)\Phi/2}
    {4\kappa_4^4g_4^4B^2\rho^2+(2\kappa_4^2B^2Z+g^2_4e^{2V}(k_1^2+k_2^2)\Phi/2)^2} \right|_{r_h} .
\end{align}

Here
\begin{equation}
    \rho = -Z e^{2V}a_h
\end{equation}
is the charge density and
\begin{equation}
    s = 4\pi e^{2V_h}
\end{equation}
is the entropy density.

One important observation for the following is that there is {\em explicit} dependence on $T$ in the above formulas
(the only explicit dependence on $T$ in $\delta h_{tx}, \delta h_{ty}$ was through the factor $T\xi$, which was factored out 
in order to obtain the coefficients $\a_{xi}, \sigma_{xi}$).

\section{Susceptibilities of the general model with perturbations}

The susceptibilities of the model are the double derivatives of the thermodynamic potential,
\begin{equation}
    \chi_{ab} = \frac{1}{\rm Vol} \left.\frac{\partial^2 \Omega}{\partial a \partial b}\right|_{\rm other\;vars.},
\end{equation}
where  $a$ and $b$ stand for the thermodynamic variables.

The potential is given by the on-shell Euclidean action times the temperature
\begin{equation}\label{eq:thermoPot}
    \Omega = T S_E\;,
\end{equation}
so we need to compute the Euclidean action 
\bea
   S_E&=& \int d^4\textbf{x} \left(
    - \frac{F_{\mu\nu}F^{\mu\nu} Z(\phi)}{4 g_4^2}
    - F_{\mu\nu}\tilde{F}^{\mu\nu} W(\phi)\right.\cr
&&\left.   R-V(\phi) + \frac{-\frac{1}{2} (\partial^\mu\phi) (\partial_\mu\phi )-\frac{1}{2} ((\partial \chi_1)^2
+(\partial \chi_2)^2) \Phi (\phi)}{16 \pi  G_N}\right)
\eea
on the ansatz (\ref{eq:ansatzIni} \ref{eq:ansatzFim}), this time up to quadratic terms in $\epsilon$. 

The integral over time cancels with the temperature in (\ref{eq:thermoPot}), and the integrals over $x$ 
and $y$ turn into an overall volume $\text{Vol}=\int dx\int dy$, so in the end our result will only depend on an integral over $r$.

\subsection{Susceptibilities with $(a,b)\in (\xi,E,B)$}

The full quadratic Lagrangian is too big for us to show in this paper, but luckily a lot of terms goes to zero 
when we take the double derivatives. Furthermore, counterterms also do not contribute at this level. 

Here we calculate and show these facts for 
the off-diagonal susceptibilities involving the magnetic $B$ and electric $E$ fields and the thermal gradient $\xi$:
\begin{align}
    \chi_{E\xi} & =  \int_{r_h}^{\Lambda} dr \left( -\frac{ a(r) Z(\phi )}{g_4^2 U(r)} \right),
    \\
    \chi_{BE} & = \int_{r_h}^{\Lambda} dr \left( -\frac{ \delta h_{ty} Z(\phi)}{g_4^2 U(r)} \right),
     \\
    \chi_{\xi B} & = \int_{r_h}^{\Lambda} dr \left(
    \frac{ a(r) \delta h_{ty} Z(\phi) }{g_4^2 U(r)} + \mathcal{O}(t)
    \right).
\end{align}

We also obtain formulas for the diagonal susceptibilities involving the same:
\begin{align}
    \chi_{EE} & =  \int_{r_h}^{\Lambda} dr \left(
    \frac{ Z(\phi)}{g_4^2 U(r)}
    \right),
    \\
    \chi_{\xi \xi} & = \int_{r_h}^{\Lambda} dr \left(
    \frac{ a(r)^2 Z(\phi)}{g_4^2 U(r)} + \mathcal{O}(t^2)
    \right),
    \\
    \chi_{BB} & = \int_{r_h}^{\Lambda} dr \left(
    \frac{Z(\phi) }{g_4^2 U(r)} \left(\frac{ \left(\delta h_{tx}^2+\delta h_{ty}^2-U(r)^2 
    \left(\delta h_{rx}^2+\delta h_{ry}^2\right)\right)}{2}- U(r) e^{-2 V(r)}\right) + \mathcal{O}(t)
    \right). 
\end{align}

\subsection{Susceptibilities with $(a,b)=(T,...)$}

We wish to also compute the susceptibilities involving the temperature $T$ as (at least) one of the variables $(a,b)$. 
One way to do this is 
to solve the integral of $r$ and get a result that depends on the fields at the boundary and at the horizon, 
while the latter is related to the temperature. This proved to be a hard challenge in this general case, since 
we obtain functions that are not calculable with the methods we employ.

Instead, the path we explored is to make use of the already computed result for the electrical currents (\ref{eq:defCurrent}), 
and consider only the case that the $T$ dependence comes only from explicit dependence, not from {\em implicit}
$T$ dependence in the conductivities $\sigma_{xx},\a_{xx}$ (previously computed) and in the metric fluctuations 
$\delta h_{tx},\delta h_{ry}$.

First, we use the fact that
\begin{equation}
    \mathcal{J}^i (r) = \mathcal{J}^i (r_h) = J^i\;,
\end{equation}
since $\mathcal{J}_i$ does not depend on $r$. Thus we can relate the fields at any $r$ through the result for 
the thermoelectric response (\ref{eq:OhmLaw}),
\begin{equation}
    J^i = \sigma_{xi} E - \alpha_{xi} T\xi\;,
\end{equation}
where we have computed $\sigma_{xi}$ and $\alpha_{xi}$ in subsection \ref{sec:Transport}, where we noted that they had 
no {\em explicit} $T$ dependence.

Then we obtain
\begin{equation}
  \sqrt{-g}\left(F^{r i} + W\tilde{F}^{ri} \right) - \epsilon \xi  M(r)\delta_{iy} 
  =
  \sigma_{xi} E - \alpha_{xi} T\xi.
\end{equation}

Solving for $\xi$ the above equation for $i=x$, we have
\begin{equation}
    \xi = \frac{Z(\phi) \left(\delta h_{tx} e^{2 V(r)} a'(r)+U(r) \left(B \delta h_{ry}
    +\delta A_x'\right)\right)+E g_4^2 \sigma_{xx}}{\alpha_{xx} g_4^2
   T}.
\end{equation}

Then we substitute $\xi$ as a function of $T$ from the above formula in the quadratic Lagrangian, 
and after taking derivatives ({\em and assuming $\delta h_{tx}, \delta h_{ry}$ and $\sigma_{xx},\a_{xx}$ 
are $T$-independent, i.e., considering only the explicit dependence in their formulas}) we have, at lowest order in $T$,
\bea
    \chi_{ET} & =&  \int_{r_h}^{\Lambda} dr 
    \frac{2 \sigma_{xx} a(r)^2 Z(\phi) }{\alpha_{xx}^2 g_4^4 T^3 U(r)} \left(Z(\phi) 
    \left(\delta h_{tx} e^{2 V(r)} a'(r)\right.\right.\cr
    &&\left.\left.+U(r) \left(B \delta h_{ry}+\delta A_x'\right)\right)+E
   g_4^2 \sigma_{xx}\right).
\eea

Rewriting this, we get the final form,
\begin{align}
    \chi_{ET} & =  \int_{r_h}^{\Lambda} dr \left(\frac{2 \sigma_{xx} a(r)^2 Z(\phi) \xi}{\alpha_{xx} g_4^2 T^2 U(r)} 
    \right).
\end{align}

We can do the same procedure to find the other susceptibilities involving $T$, at the lowest order in $T$,
\bea
    \chi_{BT} & =&  \int_{r_h}^{\Lambda} dr \;2 \epsilon ^2 a(r)^2 \delta h_{ry} Z(\phi)^2 \left(
    \frac{Z(\phi) \delta h_{tx} e^{2 V(r)} a'(r)}{\a_{xx}^2 g_4^6T^3}\right.\cr
    &&\left.+\frac{Z(\phi)U(r) \left(B \delta h_{ry}+\delta
   A_x'(r)\right)+E g_4^2 \sigma_{xx}}{\alpha_{xx}^2 g_4^6 T^3}
    \right)\cr
    \chi_{TT} & =& -\frac{T}{\rm Vol}\left.\frac{\d \Omega}{\d T^2}\right|_{B,\mu}=
     \int_{r_h}^{\Lambda} dr 3 \epsilon ^2 a(r)^2 Z(\phi)\times\cr
    &&\times\left(\frac{ \left(Z(\phi) \left(\delta h_{tx} e^{2 V(r)} a'(r)+U(r) 
    \left(B \delta h_{ry}+\delta A_x'\right)\right)+E g_4^2
   \sigma_{xx}\right)^2}{\alpha_{xx}^2 g_4^6 T^4 U(r)}
    \right).    
\eea

Note the sign difference, and the multiplication by $T$, which are standard for $\chi_{TT}$.

Rewriting these, we get
\bea
    \chi_{BT} & =&  \int_{r_h}^{\Lambda} dr \left(\frac{2 \epsilon ^2 a(r)^2 \delta h_{ry} Z(\phi)^2 \xi}{\alpha_{xx} g_4^4 T^2}
    \right)\cr
    \chi_{TT} & = & \int_{r_h}^{\Lambda} dr \left(
    \frac{3 \epsilon ^2 a(r)^2 Z(\phi) \xi^2}{\alpha_{xx} g_4^4 T^2 U(r)}
    \right).    \label{chiBTTT}
\eea

\subsection{Comparison with dyonic black hole results}

In \cite{Hartnoll:2007ai,Hartnoll:2007ih}, the thermodynamic potential $\Omega(T,\mu, B)$ was calculated for the 
$AdS_4$ dyonic black hole in the absence of the topological term $W$, obtaining 
\be
\frac{\Omega}{V}=\frac{c\a^3}{4\pi}\left(-1-\frac{\mu^2}{\a^2}+3 \frac{B^2}{\a^4}\right)\;,
\ee
where 
\be
\frac{c}{4\pi}=\frac{\sqrt{2}N^{3/2}}{6\pi}\frac{1}{4}
\ee
and $T(\a,B,\mu)$ is obtained from $\a$ from the equation 
\be
\frac{4\pi T}{\a}=3-\frac{\mu^2}{\a^2}-\frac{B^2}{\a^4}.\label{TalphaBmu}
\ee

Then the entropy density and charge density are obtained from the first derivatives of $\Omega$, 
\bea
s&=&\frac{S}{V}=-\frac{1}{V}\left.\frac{\d \Omega}{\d T}\right|_{B,\mu}=c\a^2\cr
\rho&=&-\frac{1}{V}\left.\frac{\d \Omega}{\d \mu}\right|_{B,T}=\frac{c}{\pi}\a\mu
\eea
and the matrix of susceptibilities with respect to $T$ and $\mu$ is obtained from the second derivatives \cite{Melnikov:2020ktj},
\bea
\chi_{\mu\mu}&=&-\frac{1}{V}\left.\frac{\d^2\Omega}{\d \mu^2}\right|_{B,T}=\frac{6c\a_0^3}{6\a_0^2-\mu^2}+{\cal O}(T)\cr
\chi_{TT}&=&-\frac{T}{V}\left.\frac{\d^2\Omega}{\d T^2}\right|_{B,\mu}=\frac{4c\pi \a_0^3}{6\a_0^2-\mu^2}T+{\cal O}(T^2)\cr
\chi_{\mu T}&=& -\frac{1}{V}\left.\frac{\d^2\Omega}{\d T\d \mu}\right|_B=\frac{2c\mu \a_0^2}{6\a_0^2-\mu^2}+{\cal O}(T).
\eea

Note that $\chi_{TT}$ is linear in $T$ at small $T$ (due to the multiplication by $T$ of the double derivative).

But one can consider also the topological term $W$, as was done in \cite{Nastase:2022etj}, and find the thermodynamical 
potential 
\be
\frac{\Omega}{V}=\frac{c\a^3}{4\pi}\left(-1-\frac{\mu^2}{\a^2}+3 \frac{B^2}{\a^4}+4W\frac{\mu B}{\a^3}\right)\;,
\ee
and in that case we obtain a modification in the charge density,
\be
\rho=\frac{Q}{V}=-\frac{1}{V}\left.\frac{\d \Omega}{\d \mu}\right|_{B,T}=\frac{c}{\pi}(\a\mu-WB)\;,
\ee
but not in the entropy density formula (as a function of $\a$), $s=c\a^2$. 

The magnetization density is now 
\be
M=\frac{1}{V}\left.\frac{\d \Omega}{\d B}\right|_{T,\mu}=\frac{c}{\pi}\left(\frac{B}{\a}+W\mu\right).
\ee

From (\ref{TalphaBmu}), we obtain at $T,\mu$ fixed $\a=\a(B)$, giving
\be
d\a\left(3+\frac{\mu^2}{\a^2}+3\frac{B^2}{\a^4}\right)=2B\frac{dB}{\a^3}\;,
\ee
so that finally
\be
\chi_{BB}=\frac{c}{\pi}\frac{1}{\a}\left(1-\frac{2B^2}{3\a^4+\mu^2\a^2+3B^2}\right).
\ee

Putting $T\simeq 0$ in (\ref{TalphaBmu}), we obtain 
\be
\chi_{BB}\simeq \frac{c}{\pi}\frac{2\a_0}{4\a_0^2-\mu^2}+{\cal O}(T).
\ee

For 
\be
\chi_{TB}=-\left.\frac{\d M}{\d T}\right|_{B,\mu}\;,
\ee
we obtain at fixed $B,\mu$ from  (\ref{TalphaBmu}) that
\be
dT=\frac{d\a}{4\pi}\left(3+3\frac{B^2}{\a^4}+\frac{\mu^2}{\a^2}\right)\;,
\ee
so that 
\be
\chi_{TB}=\frac{cB}{4\pi^2}\frac{\a^2}{3\a^4+3B^2+\mu^2\a^2}.
\ee

Putting $T\simeq 0$ in (\ref{TalphaBmu}), we obtain 
\be
\chi_{TB}=\frac{cB}{4\pi}\frac{1}{4\a_0^2-\mu^2}+{\cal O}(T).
\ee

We see that both $\chi_{BB}$ and $\chi_{TB}$ go to constants at $T\rightarrow 0$, while we saw that $\chi_{TT}$ was 
then linear in $T$. 

It is hard to see how this can be consistent with the formulas for $\chi_{BT}$ and $\chi_{TT}$ in (\ref{chiBTTT}), where the 
temperature appears in the denominator.
One possibility then is that our assumption of partial derivative acting only on the explicit $T$'s in the conductivities 
and metric components was wrong, but that seems unlikely. 

More likely is that, actually, the formulas derived from the $AdS_4$ dyonic solution, a "top-down" type solution, in fact do not 
match the general solution, with fields introduced as perturbations. Thus one should be very careful when importing 
results from one way of calculating into another.

\section{Conclusions}

In this work we have calculated thermodynamic susceptibilities, the second order derivatives of the thermodynamic 
potential, whose matrix is related to the conductivity matrix by the general theory of the hydrodynamic limit, for a general 
holographic model with external fields $B, B_1$ and then $E, B, \mu, \xi$ introduced as perturbations at infinity. 
In the process, we have also found more general formulas for the thermoelectric conductivities in the case that not only 
translational invariance, but isotropy is also broken, through general linear dilatons $\chi_1=k_1x, \chi_2=k_2y$, $k_1\neq k_2$.

We have then compared the formulas with formulas obtained in the standard analysis using the "top-down" $AdS_4$ 
dyonic black hole, and we have found that the results do not match. While there is a possibility that one of the assumptions 
in our calculation is unwarranted, we think that unlikely. More likely, calculations using different types of assumptions (the fields 
are nonperturbatively introduced in the dyonic black hole, while perturbatively introduced at infinity in the case considered here)
are not expected to match in general, so one should be careful when exporting them from one model to the other.

\section*{Acknowledgements}

The work of HN is supported in part by  CNPq grant 301491/2019-4 and FAPESP grants 2019/21281-4 
and 2019/13231-7. HN would also like to thank the ICTP-SAIFR for their support through FAPESP grant 2016/01343-7.
The work of CLT is supported by CNPq grant 	141016/2019-1.


\bibliography{duality2Caio}
\bibliographystyle{utphys}

\end{document}